\documentclass[fleqn,usenatbib]{mnras}

\usepackage[T1]{fontenc}

\DeclareRobustCommand{\VAN}[3]{#2}
\let\VANthebibliography\thebibliography
\def\thebibliography{\DeclareRobustCommand{\VAN}[3]{##3}\VANthebibliography}

\usepackage{graphicx}	
\usepackage{amsmath}	
\usepackage{amssymb}	
\usepackage{multirow}

\title[dynamics of the clumps from partial disruption]{Dynamics of the clumps partially disrupted from a planet around a neutron star}

\author[Kurban et al.]{
Abdusattar Kurban $^{1,2,3}$\thanks{E-mail: akurban@xao.ac.cn},
Xia Zhou $^{1,2,3}$\thanks{E-mail: zhouxia@xao.ac.cn},
Na Wang $^{1,2,3}$,
Yong-Feng Huang $^{4,5}$,
Yu-Bin Wang $^{1,6}$,
\newauthor Nurimangul Nurmamat $^{4}$
\\
$^{1}$ Xinjiang Astronomical Observatory, Chinese Academy of Sciences, Urumqi 830011, Xinjiang, China\\
$^{2}$ Key Laboratory of Radio Astronomy, Chinese Academy of Sciences, Urumqi 830011, Xinjiang, China\\
$^{3}$ Xinjiang Key Laboratory of Radio Astrophysics, Urumqi 830011, Xinjiang, China\\
$^{4}$ School of Astronomy and Space Science, Nanjing University, Nanjing 210023, China\\
$^{5}$ Key Laboratory of Modern Astronomy and Astrophysics (Nanjing University),
Ministry of Education,  Nanjing 210023, China\\
$^{6}$ University of Chinese Academy of Sciences, 19A Yuquan Road, Beijing 100049, China
}
\date{Accepted XXX. Received YYY; in original form ZZZ}

\pubyear{2023}

\begin{document}
\label{firstpage}
\pagerange{\pageref{firstpage}--\pageref{lastpage}}
\maketitle

\begin{abstract}

Tidal disruption events are common in the Universe, which may occur in
various compact star systems and could account for many astrophysical
phenomena. Depending on the separation between the central compact star and
its companion, either a full disruption or a partial disruption may occur.
The partial disruption of a rocky planet around a neutron star can produce
kilometer-sized clumps, but the main portion of the planet can survive.
The dynamical evolution of these clumps is still poorly understood.
In this study, the characteristics of partial disruption of a rocky planet
in a highly elliptical orbit around a neutron star is investigated.
The periastron of the planet is assumed to be very close to the neutron
star so that it would be partially disrupted by tidal force every time
it passes through the periastron. It is found that the fragments generated
in the process will change their orbits on a time scale of a few
orbital periods due to the combined influence of the neutron star and
the remnant planet, and will finally collide with the central neutron
star. Possible outcomes of the collisions are discussed.

\end{abstract}




\section{Introduction} \label{sec:intro}

Tidal disruption happens when an object gets too close to its
compact host.
Tidal disruption associated with black holes (BHs) has been extensively
studied (see \citet{Guillochon2013ApJ}, \citet{Ryu2020ApJ}, \citet{Gezari2021ARAA} \citet{Mageshwaran2021NewA}, \citet{Rossi2021SSRv}
and references therein). For the disruption of planetary objects,
pioneer works have been done
by \citet{Faber2005Icar}, \citet{Guillochon2011ApJ}, and \citet{Liu2013ApJ} on gas giants.
Especially, \citet{Faber2005Icar} and \citet{Liu2013ApJ}
simulated the single tidal encounter of a close gas giant,
whereas the cases of multiple passage encounters were studied
in \citet{Guillochon2011ApJ} and \citet{Veras2014MNRAS}.

Tidal disruption of minor-planets/asteroids around white dwarfs (WDs) has been
extensively studied \citep{Vanderburg2015Natur, Granvik2016Natur}.
Recent simulations \citep{Malamud2020a,Malamud2020b} show that
a planet in a highly eccentric orbit around a WD could be disrupted
by tidal force, and materials in the inner side of the orbit would
be accreted by the WD. The material accreted by the WD may be
responsible for the pollution of the WD atmosphere by heavy
elements \citep{Vanderburg2015Natur, Malamud2020a, Malamud2020b}.
Similar processes can also occur in neutron star-planet systems if the initial
parameters of the planetary system fulfill the tidal disruption condition
\citep{Geng2015ApJ_a,Huang2017ApJ,Kuerban2019AIPC,Kuerban2020ApJ,Kurban2022ApJ}.

Depending on the relative separation between the compact star and the
companion object, the degree of disruption could be very different.
Either a full disruption or a partial disruption may occur.
In full disruption, the companion is completely destroyed.
In this case, the time for the debris material to be completely
accreted by the compact star could be very long especially when no
additional forces connected to sublimation
or radiation are involved \citep{Veras2014MNRAS}.
In the case of a partial disruption, part of the material
is stripped off from the orbiting object. For example,
the partial disruption of a rocky planet can produce kilometer-sized
clumps, but the main portion of the planet can still survive \citep{Malamud2020a}.
The fate of clumps is then affected by the combined action of
the surviving planet and the compact star, the studying of which
is still lacking in the literature.

Most of the previous studies mainly focus on the tidal disruption
of an object in a parabolic or elliptic orbit. Formation of such an orbit
is considered to be the results of various dynamical processes such as
tidal capturing \citep{Goulinski2018MNRAS,Kremer2019ApJ},
Kozai-Lidov effect \citep{Lidov1962,Kozai1962,Naoz2016ARAA,Shevchenko2017ASSL},
or scattering \citep{Hong2018ApJ,Carrera2019AA}.
The planet may have multiple close encounters with the central
star in a tidal capturing process.
In the Kozai-Lidov effect, the planet
can move close to its central star and multiple encounters could occur
naturally due to eccentricity oscillation. The close
approach is gradual in this process, and partial disruption may occur.
In addition, the probability of partial disruption is usually higher
than that of full disruption \citep{Zhong2022ApJ}.
The investigation of the dynamic behavior of the clumps produced in
a partial disruption is important, which could help understand
various astrophysical phenomena such as electromagnetic transient
events and the pollution of the WD atmosphere by heavy elements.

The Kozai-Lidov mechanism \citep{Lidov1962,Kozai1962} is an efficient
theory to model gravitational interactions in multi-body systems.
The interaction in such systems is complicated, and the evolution
could be rapid, although no dissipation is involved. Merger
events could be triggered and various bursts or transients may
be produced in these multi-body systems \citep{Naoz2016ARAA,Shevchenko2017ASSL}.

In standard Kozai-Lidov theory \citep{Lidov1962, Kozai1962},
the angular momentum of the test particle is conserved when
a circular orbit is assumed for the outer perturber, causing
periodic variation of the test particle's eccentricity and inclination.
However, the angular momentum for both inner and outer orbits
are not conserved when the perturber's orbit is eccentric,
which leads to very different behaviors of the test particle
\citep{Lithwick2011ApJ, Li2014ApJ, Naoz2017AJ}.

In a three-body system, the orbit parameters significantly
influence the dynamic outcome. The hierarchy of such
a system is mainly determined by the parameter 
$ \epsilon = a_{1} e_{2}/[a_{2}(1 - e_{2}^2)]$
(the coefficient of the octupole-order secular interaction term),
or by the factor $a_{2}(1 - e_{2})/a_{1}$, both of which parameterize the
size of the external orbit as compared to the inner binary orbit.
Note that $e$ is the eccentricity and $a$ is the semi-major axis
of the inner and outer orbits, denoted by subscripts 1 and 2 respectively.
A system with $\epsilon < 0.1 $ is hierarchical and stable.
It is mildly hierarchical when $0.1 <\epsilon < 0.3 $, and is
non-hierarchial when $\epsilon > 0.3 $ \citep{Naoz2016ARAA}.
Equivalently, when the factor of $a_{2}(1 - e_{2})/a_{1}$ is
concerned, a system with $a_{2}(1 - e_{2})/a_{1} > 10$ is hierarchical.
It is moderately hierarchical when $ a_{2}(1 - e_{2})/a_{1} = $ 3---10,
and it is non-hierarchial when $a_{2}(1 - e_{2})/a_{1} < 3$
\citep[e.g.][]{Antonini2012ApJ,Katz2012,Antonini2014ApJ,He2018MNRAS}.
It has been shown that the orbit-averaging method (including quadrupole
and octupole terms) is broken down in non-hierarchial, dynamically unstable systems,
and this leads to an underestimation of the maximum eccentricity
\citep{Antonini2012ApJ, Katz2012,Antonini2014ApJ, He2018MNRAS,Grishin2018MNRAS, Bhaskar2021AJ}.
It was also found that a system with $ a_{2}(1 - e_{2})/a_{1} < 10$
follows a complex dynamical evolution pattern that
can lead to a very large eccentricity and secular approximations can not
accurately describe the evolution features.

The hierarchical triple systems in test particle limit
have been studied for different scientific cases
\citep{Lithwick2011ApJ,Katz2011PhRvL,Naoz2013MN,Li2014ApJ,
Li2014ApJ_b,Li2016ApJ,Li2018AJ}.
For a system with nearly coplanar
(the inclination $ i \sim 0 $) and highly eccentric
(for both inner and outer orbits) configuration, \citet{Li2014ApJ}
found that the eccentricity of the inner test particle
increases steadily due to the perturbation (this effect
is significant for tight configurations).
Recently, \citet{Grishin2018MNRAS} and \citet{Bhaskar2021AJ}
studied the dynamic evolution of highly inclined, mildly
hierarchical systems and found that the orbit-average method can
describe the evolution of the inner orbit well when $e_{2} = 0$.
But \citet{Bhaskar2021AJ} found that the orbit-average method is
not accurate enough for systems with $ a_{1}/a_{2} > 0.3 $ and
$m_{2}/M < 6\times10^{-3}$ (here $M$ is the central star mass), and the accuracy
is worsening with the increase of $a_{1}/a_{2}$ (i.e. for a tighter configuration)
and the decrease of $m_{2}/M$. Thus the value of $a_{2}(1 - e_{2})/a_{1}$
can effectively reflect the stability of the system.
If the configuration of the system does not
satisfy the stability condition \citep{Mardling2001MNRAS,He2018MNRAS}
or the secular approximation of the Kozai-Lidov mechanism
is broken-down \citep{Antonini2012ApJ, Antonini2014ApJ, Antonini2016ApJ},
the orbit of the test particle will evolve even faster, essentially
causing a collision or ejection.

The goal of this study is to investigate the dynamical evolution of
the clumps produced during the partial disruption of a rocky planet
that moves around a neutron star (NS) in a highly elliptical orbit.
The motion of these clumps is affected by both the NS and the
planet's surviving core, making the system an ideal case
for the Kozai-Lidov mechanism. The structure of our paper is as follows.
In Section \ref{sec:structure}, the structure of the planet is described,
which is an important factor that affects the dynamics of the clumps.
The tidal disruption process is introduced in Section \ref{sec:rtd},
paying special attention on the effects of various parameters.
Section \ref{sec:orbit-distribution} presents the
distributions of orbital parameters for the clumps.
The dynamics of the clumps are detailedly investigated in Section \ref{sec:dynamics},
and possible factors that can affect the dynamics are then described
in Section \ref{sec:factors}.
Finally, Section \ref{sec:conclusion} presents our conclusions
and some brief discussion.

\section{The structure of planet} \label{sec:structure}

Up to now, more than 5000 exoplanets have been detected. Their
measured parameters (mass, radius) indicate that their composition
should be different from each other. Various equations of states
(EOSs) are proposed for exoplanets
\citep{Benz1999Icar,Seager2007ApJ,Swift2012ApJ,Howe2014ApJ,
	Smith2018NatAs,Otegi2020AA}.
In this study, we mainly focus on rocky planets which orbit
around neutron stars. Three types of planets will be considered,
i.e. pure iron (Fe) planets, pure perovskite (MgSiO$_{3}$) planets,
and the two-layer planets composed of a Fe core and a MgSiO$_{3}$ mantle.
For Fe materials, we use the EOS derived based on experimental data,
as described in \citet{Smith2018NatAs}, which is expressed as
\begin{equation}\label{eq:eos_Fe}
P_{\rm SM} = 3K_{0} x^{-2}\left(1 -x\right) \exp\left[\left(1.5K'_{0} - 1.5\right)\left(1 - x\right)\right],
\end{equation}
where $ x = (\rho_{0}/\rho)^{1/3}$,
$ \rho_{\rm 0} = 8.43 \rm\,g\,cm^{-3}$ (the density at zero pressure),
$ K_{0} = 177.7 \rm \, GPa $ and $ K'_{0} = 5.64 $.
For MgSiO$_{3}$ materials, we take the third-order finite strain Birch-Murnagham
EOS that is widely used for exoplanet modeling \citep{Birch1947PhRv,Seager2007ApJ},
i.e.
\begin{equation}\label{eq:eos}
P_{\rm BM} = \frac{3}{2}K_{0} \left(y^{7/3} - y^{5/3}\right) \left[1 + \frac{3}{4}\left(K'_{0} - 4\right)\left( y^{2/3} - 1\right)\right],
\end{equation}
where $ y = (\rho_{0}/\rho)^{-1}$. In this case, the density at zero pressure
is $ \rho_{\rm 0} = 3.22 \rm\,g\,cm^{-3}$, and the two constants are
$ K_{0} = 125 \rm \, GPa $ and  $ K'_{0} = 5 $, respectively.

\begin{figure}
	\includegraphics[width=\columnwidth]{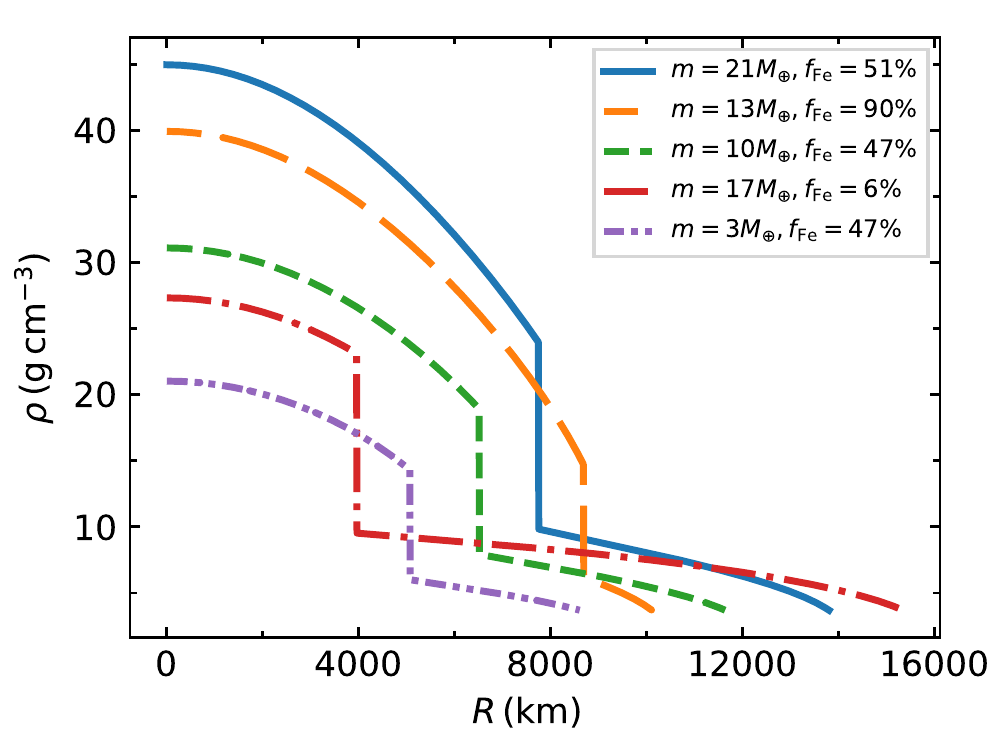}
	\caption{$\rho$ vs. $ R $ for two-layer rocky planets with different
	Fe core mass fraction $f_{\rm Fe}$.
	The EOS of \citet{Smith2018NatAs} is used for the Fe core
	and the EOS of \citet{Seager2007ApJ} is used for the
	MgSiO$_{3}$ mantle (see main text for details).
	For similar plots, see \citet{Seager2007ApJ} and \citet{Howe2014ApJ}.}
	\label{fig:r_rho}
\end{figure}

\begin{figure}
	\includegraphics[width=\columnwidth]{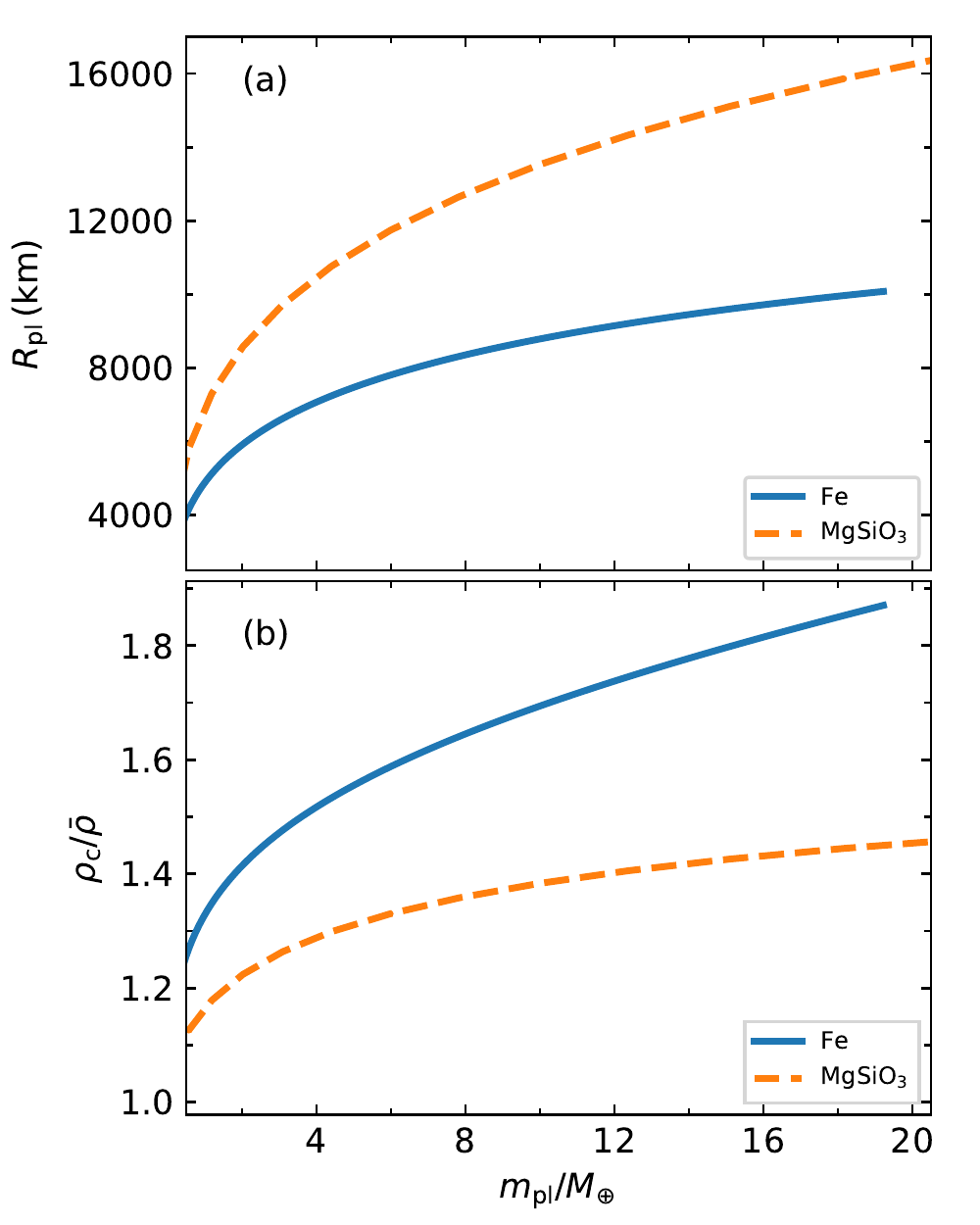}
	\caption{(a) Mass-Radius relation and (b) $ \rho_{\rm c}/\bar\rho$ vs. mass
		for pure Fe planets and pure MgSiO$_{3}$ planets.
		For similar plots, see \citet{Smith2018NatAs} and \citet{Seager2007ApJ}.}
	\label{fig:m-r}
\end{figure}

The mass, radius, and internal density profile of a planet
can be calculated by using the above EOSs.
The property of a planet composed of pure materials
is mainly characterized by its central pressure ($P_{\rm c}$). But for a
two-layer planet composed of a Fe core and a MgSiO$_{3} $ mantle,
both the central pressure and the pressure at the core-mantle boundary
($P_{\rm cb}$) are necessary to determine the mass and radius of
the planet. In our calculations, we will take various possible $P_{\rm c}$
and $P_{\rm cb}$ values \citep{Seager2007ApJ,Howe2014ApJ} to study their
effects on the dynamics. For a two-layer planet, the EOS can be
generally expressed as
\begin{equation}\label{eos_two}
	P = \left\{
	\begin{array}{lr}
		P_{\rm SM},~~~~ P_{\rm } \geq P_{\rm cb} ~~~~(\rm for~core)\\
		P_{\rm BM}.~~~~ P_{\rm } < P_{\rm cb} ~~~~ (\rm for~mantle)
	\end{array}
	\right.
\end{equation}

As an example, Fig. \ref{fig:r_rho} plots the $ \rho-R $
(density vs. radius) profile of some two-layer
planets. Note that the
$P_{\rm c}$ and $P_{\rm cb}$ are different
for these objects. On each curve, the vertical segment
represents the boundary between the core and the mantle,
at which there is a jump in density but the pressure is
still continuous, that is $P_{\rm cb}$. The mass-radius ($ m_{\rm pl}-R_{\rm pl} $)
relation of pure Fe planets and pure MgSiO$_{3}$ planets
is shown in Panel (a) of Fig. \ref{fig:m-r}, and
the relation between $m_{\rm pl}$ and $ \rho_{\rm c}/\bar{\rho}$
(the ratio of the central density to the mean density) is shown
in Panel (b). It can be seen that planets consisting of pure
Fe are more compact than those composed of MgSiO$_{3}$.

\section{Condition of tidal disruption}\label{sec:rtd}

Let us consider the tidal disruption of a rocky
planet by its host, an NS. The planet moves around the NS
in an eccentric orbit. The mass of the NS is designated as
$ M_{\star}$ (in the calculations throughout this
paper, we take $ M_{\star} = 1.4 M_{\sun}$),
and the planet is a rocky object with a mass of $ m_{\rm pl} $
and an orbital period of $ P_{\rm orb} $. According to the Kepler's
third law, the semi-major axis ($ a $) of the orbit is
related to $ P_{\rm orb} $ as
\begin{eqnarray} \label{eq:kepler_3}
	\frac{P_{\rm orb}^2}{a^3} = \frac{4 \pi^2}{G(M_{\star} + m_{\rm pl})}.
\end{eqnarray}
The separation between the planet and the central star in an
eccentric orbit is phase-dependent. At phase $ \theta $, it is
$r = a(1-e^2)/(1 + e \,\cos\theta)$,
where $ e $ is the eccentricity of the orbit. Note that
the periastron is $ r_{\rm p} = a(1 - e) $ in this case.

If the planet is too close to the NS, the central star's tidal
force would exceed the planet's self-gravity at the surface so that
the planet will be disrupted by the tidal force. The critical separation
is defined as the tidal disruption radius \citep{Hills1975Natur}.
For a gravity-dominated object, the tidal disruption radius can be
expressed as
\begin{eqnarray} \label{eq:r_td1}
	r_{\rm td} = R_{\rm pl}\left( \frac{2M_{\star}}{m_{\rm pl}}\right)^{1/3}.
\end{eqnarray}
For a rocky planet, the tidal disruption radius is $r_{\rm td} \sim 10^{11}$ cm.
On the other hand, when the distance of the planet
is only slightly larger than $r_{\rm td}$, it will also be affected by
the tidal force and could be partially disrupted. Fragments of several
kilometers in size will be produced during this process.
The degree of partial disruption depends on the separation ($ r $).
The details relevant to partial disruption will be discussed in the
next section.

Small bodies with a radius of less than a few kilometers are bounded by
material-strength. For them, the intrinsic material shear and cohesive
strengths can help resist the tidal
force \citep{Sridhar1992Icar,Holsapple2008Icar,Zhang2020NatAs}.
According to the elastic-plastic continuum theory \citep{Holsapple2008Icar},
the tidal disruption limit of small bodies in the
the material-strength dominated regime is \citep{Zhang2020NatAs,Zhang2021ApJ}	
\begin{eqnarray} \label{eq:td_for-smallbody}
	r_{\rm str} = \left( \frac{\sqrt{3}M_{\star}}{4\pi\rho_{\rm cl}}\right)^{1/3}
	\left(\frac{5k}{4G\pi\rho_{\rm cl}^2 r_{\rm cl}^2} + s \right)^{-1/3},
\end{eqnarray}
where $ r_{\rm cl} $ and $ \rho_{\rm cl} $ are the radius and density
of the small body (clump), respectively. The two constants
$ s $ and $ k $ can be expressed as a function of the friction angle
$\phi$ and the cohesive strength $ C $, $ s = 2 \sin\phi/\sqrt{3}(3 -\sin\phi)$
and $ k = 6C \cos\phi/\sqrt{3}(3 -\sin\phi)$. The friction angle usually
ranges in $25^{\circ}$ --- $50^{\circ}$ \citep{Holsapple2008Icar,Bareither2008}.
The cohesive strength of meteorites/asteroids is typically in the range of
0.1-10 MPa \citep{Pohl2020MPS,Veras2020MNRAS},
while it is $ \sim $1 Pa for comets \citep{Gundlach2016AA}.
In our case, the small clump typically has a size of 2 km and a density of
$ \rho_{\rm cl} = 5 \,\rm g\,cm^{-3}$ (typical density for
MgSiO$_{3}$). Taking $\phi = 30^{\circ}$, then the break up distance is
$r_{\rm str} = 5.31\times10^{10}$ cm for $ C $ = 0.1 MPa, and it is
$r_{\rm str} = 9.88\times10^{9}$ cm for $ C $ = 10 MPa.
Alternatively, taking $\phi = 45^{\circ}$, we have
$r_{\rm str} = 2.41\times10^{10}$ cm for $ C $ = 0.1 MPa, and
$r_{\rm str} = 5.34\times10^{9}$ cm  for $ C $ = 10 MPa.
It means that the small body will remain intact even when they are very
close to the central star.


\section{Orbits of the clumps}\label{sec:orbit-distribution}


In this study, we consider a rocky planet moving around an NS
in a highly elliptical orbit. The periastron is assumed to be
slightly larger than the tidal disruption radius so that the
planet will be partially disrupted every time it passes through
the periastron. A few clumps of several kilometers are
generated in the process. The condition of the NS-planet system
determines the features of the stripped clumps.
First, the degree of disruption is determined by the ratio of the
tidal disruption radius with respect to the periastron, which we
define as the penetration factor ($\beta$), i.e. $ \beta = r_{\rm td} / r_{\rm p} $.
If $\beta $ is large (the separation is small at the periastron),
then the disruption is violent and the planet would be significantly
destroyed. On the other hand, when $ \beta $ is small (the
separation is large), the planet will only be lightly affected. The
material in the outer crust of the planet will be stripped off to
produce some rocky clumps, but the planet itself will largely
remain its integrity
\citep[e.g.][]{Manser2019Sci,Guillochon2011ApJ,Liu2013ApJ, Malamud2020a,
	Malamud2020b,Law-Smith2020ApJ}.

The mass stripped from the planet depends on its structure
and the penetration factor ($\beta$). The mass loss
($ \Delta m $) is very small when $\beta \lesssim 0.5$
\citep{Guillochon2011ApJ,Liu2013ApJ,Ryu2020ApJ,Law-Smith2020ApJ}.
It corresponds to a periastron separation of
$r_{\rm p} \gtrsim 2r_{\rm td}$. Here we take $r_{\rm p} =
2r_{\rm td}$ as a typical condition for partial disruption,
which is used to determine the orbit parameters of the planet
and the clumps in the following calculations.

After disruption, the clumps are still bound to the central star, but
they should have slightly different orbital parameters since they
originate from different parts of the planet. It is very similar to
the disruption of a planet around
a WD \citep{Malamud2020a,Malamud2020b,Brouwers2022MNRAS,Brouwers2023MNRAS}. For example, the
semi-major axis ($a_{\rm cl}$) of the clump orbit depends on the
$\beta$ and displacement ($ R $) of the clump relative to the mass
center of the planet at the moment of breakup
($ R = 0 $ corresponds to the center of the planet).
As a result, the semi-major axis of a clump stripped off
from the inner and outer side of the planet is \citep{Malamud2020a}
\begin{equation}\label{a_cl}
a_{\rm cl} = \left\{
\begin{array}{lr}
    a \left( 1 + a\frac{2R}{d(d - R)}\right)^{-1},~~(\rm inner)\\
    a \left( 1 - a\frac{2R}{d(d + R)}\right)^{-1},~~(\rm outer)
\end{array}
\right.
\end{equation}
where $ d $ is the distance between the NS and the planet at the
moment of break up (here $ d = r_{\rm p}$).

According to Kepler's third law, the orbit
period of the clump is
\begin{equation}
    P_{\rm orb}^{\rm cl} = \left( \frac{4 \pi^2 a_{\rm cl}^3}{G (M_{\star}
		+ m_{\rm cl})}\right)^{1/2}.	
\end{equation}
As the clump returns to the periastron
of its orbit, one has $ r_{\rm p} = a_{\rm cl}(1 - e_{\rm cl}) \pm R $
(here $+$ and $-$ for inner and outer clumps), and the eccentricity is
\begin{equation}\label{e_cl}
e_{\rm cl} = \left\{
\begin{array}{lr}
    1 - \frac{r_{\rm p} - R}{a_{\rm cl}},~~(\rm inner)\\
    1 - \frac{r_{\rm p} + R}{a_{\rm cl}},~~(\rm outer)
\end{array}
\right.
\end{equation}

The spread of orbital parameters of clumps is determined by
their original locations on the planet ($ R $, inside the planet), and
by the $\beta$ parameter. Different $\beta $ leads to different outcomes,
such as a full disruption or a partial disruption.
In this study, as explained above, we mainly consider the partial
disruption cases, for which we take a typical condition of
$\beta = 0.5$ (or $ d = r_{\rm p} = 2 r_{\rm td}$).
The clumps are assumed to originate from the planet's surface ($ R \sim R_{\rm pl} $),
while the main portion of the planet remains unaffected.
As a result, the orbital parameters of the clumps, such as $a_{\rm cl}$ and $e_{\rm cl}$,
will be slightly different from that of the original planet.
For example, if we take $P_{\rm orb} = 100$ day, $m_{\rm pl} = 12.84\, M_{\oplus}$,
$R_{\rm pl} = 10107.08$ km for a two-layer planet,
and assume that materials with $\Delta R = 200 $ km in depth
are stripped off from the surface to form clumps,
i.e $R = [R_{\rm pl} - \Delta R, R_{\rm pl}]$,
then the eccentricity of the resultant clumps will be in a narrow range
of $e_{\rm cl} = [0.96469,0.96422]$. We see that for two clumps originated from
$R = R_{\rm pl} - \Delta R$ and from $R_{\rm pl}$, their difference in the
orbital eccentricity is very small. Therefore, we only need to consider the
case of the clumps in the innermost orbit.

\begin{figure*}
	\includegraphics[width=0.80\textwidth]{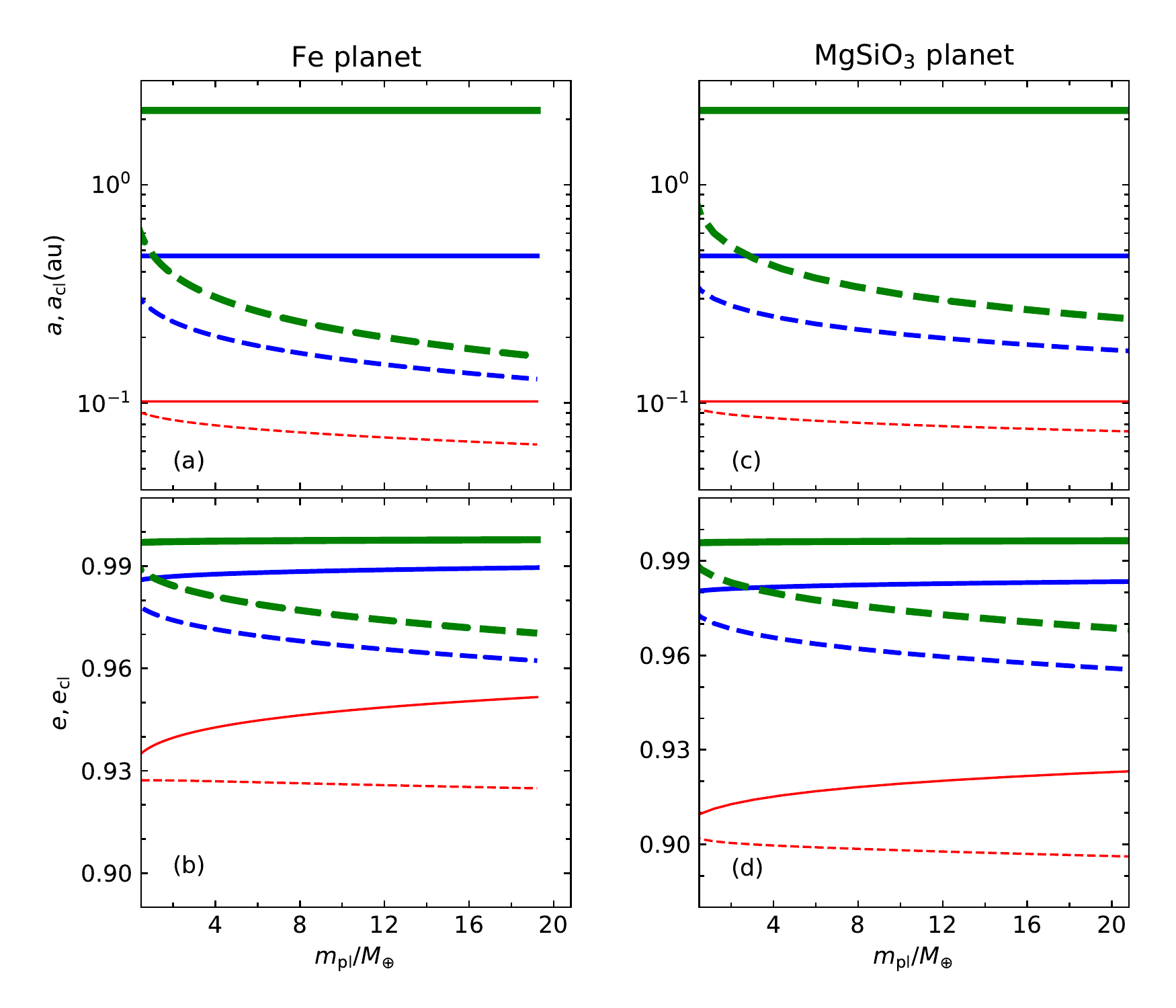}
	\caption{Panel (a) shows the planet's semi-major axis (solid lines) and the clump's
	semi-major axis (dashed lines) as a function of the planet's mass.
	The red, blue, and green lines represent the planet with an orbital
	period of 10 days, 100 days, and 1000 days, respectively.
	Panel (b) shows the planet's eccentricity (solid lines) and clump's
	eccentricity (dashed lines) as a function of the planet's mass.
	The line colors are the same as in Panel (a).
	The right panels (c) and (d) show the cases for MgSiO$_{3}$ planets,
	and the line styles are the same as that in the left panels.
	\label{fig:a-e-m1}}
\end{figure*}

We now further estimate the orbit parameters of our systems
under the partial disruption condition. Taking the planet's orbit period as
$P_{\rm orb} = 10$ days, $ 100 $ days, and $ 1000 $ days, respectively,
we have calculated the parameters of the planet and the clumps.
They are plotted versus the planet mass in Fig. \ref{fig:a-e-m1},
some exemplar parameters are presented in Table \ref{tab:table1}.
Here, the left panels show the cases for Fe planets and the right panels
show the cases for MgSiO$_{3}$ planets. We see that the planet's
eccentricity increases while the semi-major axis and eccentricity of the
clumps decrease as the planet's mass increases. At the same time,
to acquire a long orbital period, the semi-major axis and eccentricity
of the planet should be large enough.
This is easy to understand. The dependence of $a_{\rm cl}$ and
$e_{\rm cl}$ on the two parameters of $m_{\rm pl}$ and $R_{\rm pl}$ is
expressed in Equations (\ref{a_cl}) and (\ref{e_cl}). Note that in these
two equations, we have $ d = r_{\rm p} = 2 r_{\rm td}$.
From Equation (\ref{a_cl}), we see that $a_{\rm cl}$ is mainly determined
by $R$ and $r_{\rm td}$, but is almost independent
on $m_{\rm pl}$ since $m_{\rm pl} \ll M_{\star}$.
The dependence of $e_{\rm cl}$ on $m_{\rm pl}$ and $R_{\rm pl}$ is similar.

Whether the clumps are bound to the NS or not is an
important issue concerning the tidal disruption process.
Clumps in the inner stream with $ a_{\rm cl} < a$ are bound to NS.
However, for the clumps with $ a_{\rm cl} > a $ in the outer stream,
we need to consider a critical displacement of $ R_{\rm crit} = d^{2}/(2a - d) $
on the far side of the NS \citep[e.g.][]{Malamud2020a}.
The clumps coming from $ R < R_{\rm crit}$
are bound while the clumps with $ R > R_{\rm crit} $ form
parabolic orbits and are finally unbound to the NS. We will further 
investigate this issue in the next section. 

The inclination angle ($i$) of the clump is determined by its
vertical position at the surface of the planet.
The maximum inclination is $ i = arc\,\,tan(R_{\rm pl}/r_{\rm td})$.
But in typical cases, we can assume that the clump originates
from a vertical height of $R_{\rm pl}/2$, then the inclination is
$ i = arc\,\, tan(R_{\rm pl}/2 r_{\rm p})$.
For example, for a 10$ M_{\oplus} $ Fe planet, the inclination is $ i = 0.2^{\circ} $,
while for a 18$ M_{\oplus} $ Fe planet, it is $ i = 0.24^{\circ} $.
We see that the inclination is generally very small so that the clumps are
essentially coplanar with the planet.
Note that the rotation of the planet is also an important factor that can
influence the inclination of the clumps. If the planet's rotation axis is
misaligned with its orbital rotation axis, then the inclination could be
slightly larger.

\section{Dynamics of the clumps}\label{sec:dynamics}

In the above section, we have presented the condition of
partial disruption for a rocky planet around an NS.
The clumps generated during the partial disruption will
orbit around the NS under the influence of gravitational
perturbation of the surviving portion of the planet.
As a result, the orbit of each clump will evolve and the
eccentricity would increase, which will finally make
the clumps fall toward the NS. In this section, we present
details of the dynamics.

After disruption, the NS, the surviving portion of
the planet and the clumps form a multi-body system.
Ignoring the interaction between clumps,
the system can be simplified as a triple system, which
includes the NS, the remnant planet, and a particular clump.
The stability of a triple system has been studied
for a long time \citep{Eggleton1995ApJ,Mardling2001MNRAS}.
Recently, \citet{He2018MNRAS} introduced a stability
criterion for triple systems, which can be expressed as
\begin{eqnarray}\label{eq:stab}
	\frac{a\left( 1 - e \right) }{a_{\rm cl}\left( 1 + e_{\rm cl} \right) } >
	2.8\frac{1}{1 + e_{\rm cl}} \nonumber \\\times
	\left[ \left(1+\frac{m_{\rm pl}}{M_{\star}+m_{\rm cl}}\right)
	\frac{1 + e}{(1 - e)^{1/2}}\right]^{2/5} 
	\left( 1-\frac{0.3i}{180^{\circ}}\right),
\end{eqnarray}
where $ e $ and $ a $ are the orbital
parameters of the planet, $ e_{\rm cl}$ and $a_{\rm cl}$ are
the parameters of the clump.
Note that this criterion applies to any triple system.
If a system satisfies the above criteria, then it is stable.
On the contrary, if the system does not satisfy the above criteria,
then the secular evolution approximation underlying
the Kozai-Lidov theory will break down.
In such an unstable system, the clump
will either be ejected or collide with the central object.

Besides, there is another instability
that can make the orbit unstable, in which the secular approximation
is breakdown \citep{Antonini2012ApJ,Antonini2014ApJ,Antonini2016ApJ,Hamers2018MNRAS,Hamers2022ApJ}.
In this case, the evolutionary timescale of the clump's
angular momentum is comparable to the orbital period of the planet.
Several methods have been proposed to deal with the breakdown of the
secular approximation, including quadrupole approximation or even
higher order contribution \citep{Naoz2011Natur,Katz2012,Seto2013PhRvL,Naoz2013MN},
or the double-averaging approaching \citep{Katz2012,Antonini2014ApJ,
	Antonini2016ApJ,Bode2014MNRAS}. It is found that the orbit of the
clump can have a high eccentricity irrespective of its initial inclination.
As a result, the clump will finally collide with the central star
\citep[e.g.][]{Perets2012ApJ,Katz2012,
	Bode2014MNRAS,Petrovich2015,He2018MNRAS,Toonen2022AA}.

\begin{figure}
	\includegraphics[width=\columnwidth]{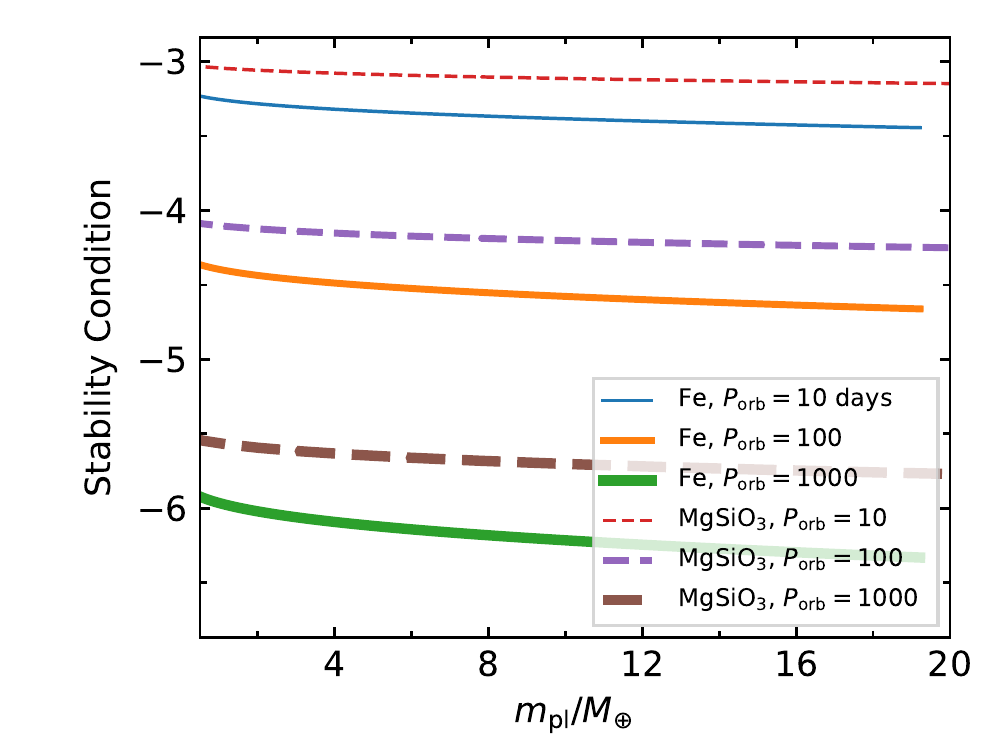}
	\caption{Stability condition for the system composed of the NS,
		the surviving portion of the planet, and the clump.
		The solid and dashed lines represent Fe and MgSiO$_{3}$ planets with
		different orbital periods,
		respectively.
		A stability condition smaller than zero means that the system is unstable.}
	\label{fig:stab}
\end{figure}

\begin{figure}
	\includegraphics[width=\columnwidth]{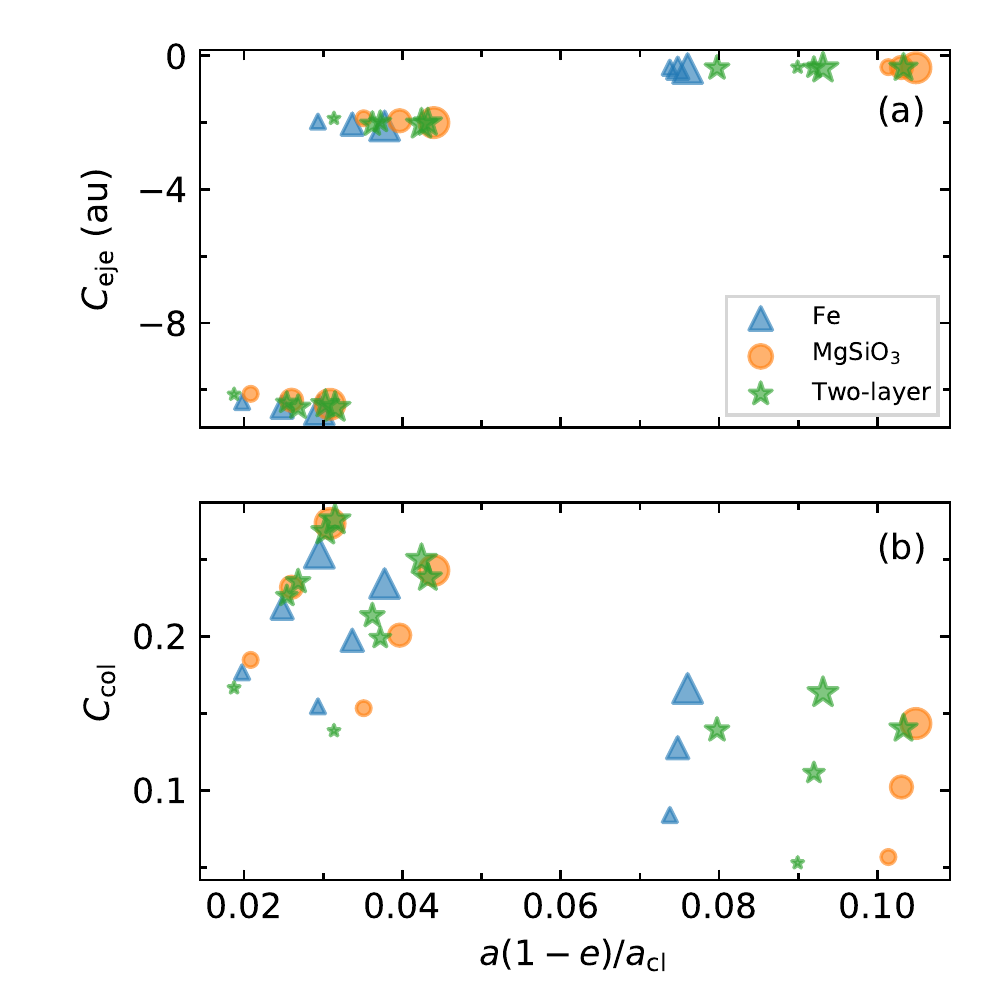}
	\caption{$C_{\rm eje}$ (a) and $ C_{\rm col}$ (b) as a function
        of $a(1 - e)/a_{\rm cl}$ for the configurations listed in
        Tables \ref{tab:table1} and \ref{tab:table2}.
        The symbol size represents the scale of the planet's mass.
        $C_{\rm eje} < 0 $ means the clump will not be ejected from
        the system. $ C_{\rm col} > 0 $ means that the clump would
        collide with NS due to the perturbation.
  } \label{fig:col-eje}
\end{figure}

We have assessed the stability of the clump's orbit by using
equation (\ref{eq:stab}). The results are plotted in Fig. \ref{fig:stab}.
Note that the stability condition is calculated by subtracting
the right-hand side from the left-hand side of Equation (\ref{eq:stab}).
It could be seen that, in our cases, the stability condition for all
configurations is smaller than zero, meaning that the orbits of the
clumps are unstable.
As stated above, such unstable clumps will either be ejected from
the system or collide with the NS.  If the separation ($r$)
between the NS and clump is larger than a critical separation of $ 5 a $,
the clump will essentially become free \citep[e.g.][]{He2018MNRAS}.
We have checked the fate of the clumps by using this method.
Since the NS has a strong magnetic field, the clumps can be magnetically
captured by the NS when they approach the magnetosphere \citep{Geng2020}.
As a result, the periastron distance of the clump's orbit should be
larger than the magnetosphere radius, which is roughly
$d_{\rm c} \sim 10^{9}$ cm. Correspondingly, the critical eccentricity
should be $e_{\rm c} = 1 - d_{\rm c}/a_{\rm cl}$, and the apocenter
distance (the largest separation between the NS and the clump on the orbit)
is $r_{\rm a,c} = a_{\rm cl}(1 + e_{\rm c})$. We then can compare
$r_{\rm a,c}$ with $5 a$ to judge the fate of the clumps. To do so,
let us define $C_{\rm eje} = r_{\rm a,c} - 5 a $. If $C_{\rm eje} > 0 $,
the clump will be ejected. Using this approach, we have checked our cases
and plotted the results in Panel (a) of Fig. \ref{fig:col-eje}.
It could be seen that none of the clumps in the inner orbit will be ejected.
This result is reasonable. The parent object of the clumps,
i.e. the planet itself, is a bound object in our framework. So, the clumps will
also be bound. On the other hand, if the planet is a free object which happens to
pass by the NS, then the disrupted clumps would mainly be unbound and would be
ejected.

On the other hand, it has been argued that the torque exerted by
the outer perturber on the clump is large enough to lead the clumps to
collide with the NS in one orbit period if the condition
\begin{eqnarray}\label{eq:stab1}
 C_{\rm col} = 2.55 \left(\frac{a_{\rm cl}}{2d_{\rm c}}\right)^{1/6} \left( \frac{m_{\rm pl}}{M_{\star}+m_{\rm cl}}\right)^{1/3} - \frac{a\left( 1 - e \right) }{a_{\rm cl}} > 0
\end{eqnarray}
is satisfied \citep[e.g.][]{He2018MNRAS}.
Using this criterion, we have checked the clumps in the system and
plotted the results in Panel (b) of Fig. \ref{fig:col-eje}.
From this figure, we could further see that $a(1 - e)/a_{\rm cl} < 0.12$
and $\epsilon > 0.3 $ are satisfied in all the cases, indicating
that they are non-hierarchical and unstable so that the secular
approximation is broken down.

In short, our system satisfies the collision condition. This is due to
the fact that the action of the surviving planet can significantly
change the clump's angular momentum at periapsis \citep[e.g.][]{Antonini2014ApJ,Zhang2023arXiv230108271Z},
leading to a quick reduction of the pericenter separation so that a head-on
collision will finally occur. During the process, the closeness of the clump's
orbit with respect to that of the surviving planet, together with the
asymmetric configuration (i.e. the eccentric orbit of the planet), is
the key factor that takes effect. Note that the perturbation
effect of the planet is strong in less hierarchical and
dynamically unstable systems \citep[e.g.][]{Toonen2022AA}.
Below, we will calculate the evolutionary timescale of the clump's
angular momentum.

The angular momentum of the clump is
$J_{\rm cl} = M_{\star} m_{\rm cl}\left( {G a_{\rm cl}(1-e_{\rm cl}^2)}/{M_{b}}\right)^{1/2}$,
where $ M_{\rm b} = M_{\star} + m_{\rm cl} $ is the total mass of the NS and the clump.
In our case, $ M_{\star}$ is much larger than $m_{\rm cl}$, so we have $M_{\rm b} \approx M_{\star} $.
Let us define $  J_{\rm cl,c} = J_{\rm cl}/(1 - e_{\rm cl}^2)^{1/2} $, which is the angular
momentum of a circular orbit with the same semi-major axis, $ a_{\rm cl} $,
then the dimensionless angular momentum of the clump   is
$ j_{\rm cl} = J_{\rm cl}/J_{\rm cl,c} = \sqrt{1 - e_{\rm cl}^{2}} $.
Due to the influence of the surviving planet, the clump's specific angular momentum
evolves on a timescale of
$t_{\rm evo} = \left( \frac{1}{j_{\rm cl}}\frac{dj_{\rm cl}}{dt}\right)^{-1}$
\citep[e.g.][]{Antonini2014ApJ,Bode2014MNRAS,Hamers2022ApJ}.
In our framework, the timescale can be further expressed as
\begin{equation}\label{eq:evo}
	t_{\rm evo} = \left(\frac{1}{j_{\rm cl}}\frac{dj_{\rm cl}}{dt}\right)^{-1} \approx
	P_{\rm orb}^{\rm cl}\frac{1}{5\pi} \frac{M_{\star}}{m_{\rm pl}}
	\left[ \frac{a(1 - e)}{a_{\rm cl}} \right]^3 \sqrt{1 - e_{\rm cl}}.
\end{equation}
It gives the time for the clump to evolve to $ j_{\rm cl} \sim 0 $,
i.e. falling toward the NS.

\begin{table*}
	\caption{Typical parameters of the Fe/MgSiO$_{3}$ planet and the clumps stripped off from the planet. \label{tab:table1}}
	\begin{tabular}{lccclcc c cccc}
		\hline
		\multicolumn{7}{c}{Planet} & &\multicolumn{4}{c}{Clump} \\
		\cline{1-7}
		\cline{9-12}
		$ \rho_{\rm c}$ &
		$ m_{\rm pl} $&
		$ R_{\rm pl} $ &
		$ \rho_{\rm c}/\bar\rho$ &
		$ P_{\rm orb} $ &
		$ a $&
		$ e $&
		&
		$ a_{\rm cl} $ &
		$ P_{\rm orb}^{\rm cl} $ &
		$ e_{\rm cl} $ &
		$ t_{\rm evo} $
		\\
		
		$(\rm g\,cm^{-3})$ &
		($M_{\oplus}$) & 
		($R_{\oplus}$) & 
		&
		(days) &
		(au) &
		&
		&
		(au) &
		(days) &
		&
		(days)\\
		\hline
		\multicolumn{12}{c}{Fe planet}\\
		\cline{1-12}
		25.65 & 4.57  & 1.15  & 1.54  & 10    & 0.10  & 0.943 &       & 0.08  & 6.73  & 0.927 & 4.74 \\
		&       &       &       & 100   & 0.47  & 0.988 &       & 0.20  & 26.83 & 0.971 & 0.75 \\
		&       &       &       & 1000  & 2.19  & 0.997 &       & 0.29  & 48.52 & 0.980 & 0.34 \\
		35.72 & 10.12 & 1.38  & 1.70  & 10    & 0.10  & 0.948 &       & 0.07  & 5.86  & 0.926 & 1.96 \\
		&       &       &       & 100   & 0.47  & 0.989 &       & 0.16  & 19.41 & 0.967 & 0.40 \\
		&       &       &       & 1000  & 2.19  & 0.998 &       & 0.21  & 30.68 & 0.975 & 0.22 \\
		48.32 & 18.12 & 1.56  & 1.85  & 10    & 0.10  & 0.951 &       & 0.07  & 5.14  & 0.925 & 1.01 \\
		&       &       &       & 100   & 0.47  & 0.989 &       & 0.13  & 14.68 & 0.963 & 0.25 \\
		&       &       &       & 1000  & 2.19  & 0.998 &       & 0.17  & 21.25 & 0.971 & 0.15 \\
		\cline{1-12}
		\multicolumn{12}{c}{MgSiO$_{3}$ planet}\\
		\cline{1-12}
		6.53  & 4.41  & 1.69  & 1.30  & 10    & 0.10  & 0.916 &       & 0.08  & 7.60  & 0.899 & 16.93 \\
		&       &       &       & 100   & 0.47  & 0.982 &       & 0.24  & 37.30 & 0.965 & 2.03 \\
		&       &       &       & 1000  & 2.19  & 0.996 &       & 0.41  & 81.58 & 0.979 & 0.72 \\
		7.95  & 9.95  & 2.12  & 1.38  & 10    & 0.10  & 0.919 &       & 0.08  & 6.94  & 0.898 & 7.23 \\
		&       &       &       & 100   & 0.47  & 0.983 &       & 0.21  & 29.07 & 0.961 & 1.07 \\
		&       &       &       & 1000  & 2.19  & 0.996 &       & 0.32  & 54.74 & 0.974 & 0.46 \\
		9.36  & 18.15 & 2.49  & 1.44  & 10    & 0.10  & 0.922 &       & 0.08  & 6.37  & 0.897 & 3.86 \\
		&       &       &       & 100   & 0.47  & 0.983 &       & 0.18  & 23.43 & 0.957 & 0.68 \\
		&       &       &       & 1000  & 2.19  & 0.996 &       & 0.26  & 39.86 & 0.970 & 0.33 \\
		\hline
	\end{tabular}
	\begin{minipage}{0.85\textwidth}
		{Note: The mass, radius, and density of the planets
			are taken from the data in Fig. \ref{fig:m-r}.
			Other parameters of the planets and clumps are calculated
			according to the equations given in the main text.}	
	\end{minipage}
\end{table*}

\begin{figure*}
	\includegraphics[width=0.80\textwidth]{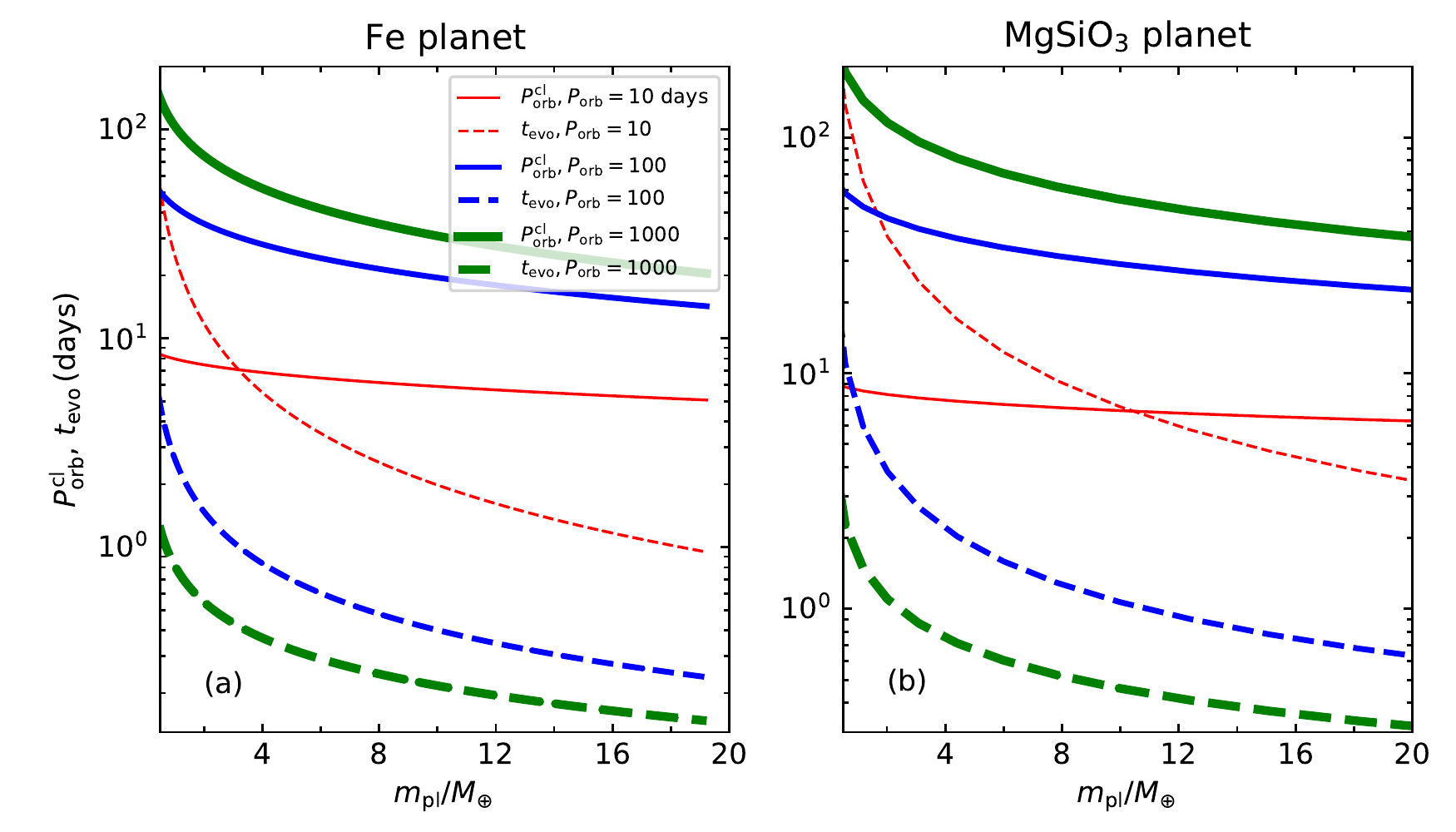}
	\caption{The orbital period of clumps (solid lines) and
	the evolution timescale of their angular momentum (dashed lines)
	plotted versus the planet's mass.
	Panel (a) is for Fe planets and Panel (b) is for MgSiO$_{3}$ planets.
	The red, blue, and green lines correspond to the planets with orbital
	period of 10 days, 100 days, and 1000 days, respectively.
	\label{fig:time}}
\end{figure*}

We have calculated the clump's orbital period ($P_{\rm orb}^{\rm cl}$) and
the corresponding evolution timescale of its angular momentum ($t_{\rm evo}$).
They are plotted versus the planet mass in Fig. \ref{fig:time}, and some
typical parameters are also listed in Table \ref{tab:table1}.
Our calculations are conducted for planets with three different orbital periods,
i.e. $P_{\rm orb} = 10$ days, $ 100 $ days, and $ 1000 $ days.
From panel (a) of Fig. \ref{fig:time}, we can see that for an
Fe planet with a mass of
$ 4.57 M_{\oplus} $, the orbital period of the clump is $6.73$ days and the evolution
timescale of its angular momentum is $4.74$ days for $P_{\rm orb} = 10$ days.
For $P_{\rm orb} = 1000$ days, the orbital period of
the clump is $48.52$ days and the evolutionary timescale of its
angular momentum is $0.34$ days.
For a Fe planet with $P_{\rm orb} = 100$ days, the orbital period of the clump
is $26.83$ days and the evolutionary timescale of its angular momentum is
$0.75$ days for $ m_{\rm pl} = 4.57 M_{\oplus} $. For $m_{\rm pl} = 18.12 M_{\oplus} $,
the orbital period of the clump is $14.68$ days and the evolutionary timescale of
its angular momentum is $0.25$ days. In general,
for a planet with a particular mass, with the increase of the planet's
orbital period, the period of the clump also increases while the
evolution timescale of its angular momentum decreases.
On the other hand, for a planet with a particular orbital period,
with the increase of the planet's mass, both the clump's orbital period and
evolution timescale of angular momentum decrease.
Panel (b) of Fig. \ref{fig:time} shows the cases for MgSiO$_{3}$ planets,
which are largely similar to that of Fe planets.

\begin{table*}
	\caption{Typical parameters of two-layer planets (with Fe core and MgSiO$_{3}$ mantle)
		and the clumps stripped off from the planet. \label{tab:table2}}
	\begin{tabular}{lcccclcc c cccc}
		\hline
		\multicolumn{8}{c}{Planet} & &\multicolumn{4}{c}{Clump} \\
		\cline{1-8}
		\cline{10-13}
		$ \rho_{\rm c}$ &
		$ m_{\rm pl} $&
		$ R_{\rm pl} $ &
		$ \rho_{\rm c}/\bar\rho$ &
		$ f_{\rm Fe} $ &
		$ P_{\rm orb} $ &
		$ a $&
		$ e $&
		&
		$ a_{\rm cl} $ &
		$ P_{\rm orb}^{\rm cl} $ &
		$ e_{\rm cl} $ &
		$ t_{\rm evo} $
		\\
		
		$(\rm g\,cm^{-3})$ &
		($M_{\oplus}$) & 
		($R_{\oplus}$) & 
		&
		(\%) &
		(days) &
		(au) &
		&
		&
		(au) &
		(days) &
		&
		(days)\\
		\hline
		21.03 & 3.26  & 1.35  & 2.89  & 47    & 10    & 0.10  & 0.925 &       & 0.08  & 7.56  & 0.911 & 14.98 \\
		&       &       &       &       & 100   & 0.47  & 0.984 &       & 0.24  & 36.71 & 0.969 & 1.82 \\
		&       &       &       &       & 1000  & 2.19  & 0.997 &       & 0.40  & 79.43 & 0.981 & 0.65 \\
		31.10 & 9.83  & 1.83  & 3.50  & 47    & 10    & 0.10  & 0.930 &       & 0.08  & 6.61  & 0.909 & 4.69 \\
		&       &       &       &       & 100   & 0.47  & 0.985 &       & 0.19  & 25.70 & 0.963 & 0.77 \\
		&       &       &       &       & 1000  & 2.19  & 0.997 &       & 0.28  & 45.54 & 0.975 & 0.36 \\
		39.92 & 12.84 & 1.59  & 2.25  & 90    & 10    & 0.10  & 0.945 &       & 0.07  & 5.80  & 0.921 & 1.91 \\
		&       &       &       &       & 100   & 0.47  & 0.988 &       & 0.16  & 18.97 & 0.964 & 0.39 \\
		&       &       &       &       & 1000  & 2.19  & 0.997 &       & 0.21  & 29.73 & 0.973 & 0.22 \\
		27.33 & 17.10 & 2.41  & 4.05  & 6     & 10    & 0.10  & 0.923 &       & 0.08  & 6.38  & 0.898 & 3.90 \\
		&       &       &       &       & 100   & 0.47  & 0.984 &       & 0.18  & 23.58 & 0.957 & 0.68 \\
		&       &       &       &       & 1000  & 2.19  & 0.996 &       & 0.26  & 40.22 & 0.970 & 0.33 \\
		44.96 & 20.68 & 2.17  & 4.01  & 51    & 10    & 0.10  & 0.935 &       & 0.07  & 5.78  & 0.908 & 2.03 \\
		&       &       &       &       & 100   & 0.47  & 0.986 &       & 0.15  & 18.81 & 0.958 & 0.42 \\
		&       &       &       &       & 1000  & 2.19  & 0.997 &       & 0.21  & 29.40 & 0.969 & 0.23 \\
		\hline
	\end{tabular}
	\begin{minipage}{0.85\textwidth}
		{Note: The mass, radius,
			and density of the two-layer planets are taken from the data in Fig. \ref{fig:r_rho}.
			Other parameters of the planets and clumps are calculated
			according to the equations given in the main text.}	
	\end{minipage}
\end{table*}

Table \ref{tab:table2} presents some typical parameters for
two-layer (Fe core and MgSiO$_{3}$ mantle) planets with different Fe-core
fraction, $f_{\rm Fe}$.
We see that with the increase in the planet's orbital period,
the period of the clump also increases, while the evolution timescale
of its angular momentum decreases, which is similar to
that of pure Fe planets and pure MgSiO$_{3}$ planets.
However, for a fixed orbital period, when the mass of the planet
increases, the changes of both the clump's period and the evolution
timescale of its angular momentum are quite different from that of
pure Fe planets and pure MgSiO$_{3}$ planets. For example, in the case
of $P_{\rm orb} = 10$ days, the clump's orbital period is $5.8$ days and
the evolution timescale of its angular momentum is $1.91$ days
for a planet mass of $ 12.84 M_{\oplus} $ ($f_{\rm Fe}=90\%$).
But for a planet mass of $ 17.10 M_{\oplus} $ ($f_{\rm Fe}=6\%$),
the clump's orbital period is $6.38$ days and the evolution timescale
of its angular momentum is $3.9$ days. It indicates that the
tidal disruption process is affected by the composition of the planets.

We now estimate the clump's travel time from the planet to
the central NS. In our framework, partial disruption
mainly occurs at the periastron. The time for a clump to
return to the periastron ($ t_{\rm ret}$) should equal to
its orbital period \citep{Zanazzi2020MNRAS,Mageshwaran2021NewA,Rossi2021SSRv}, i.e.
$ t_{\rm ret} \sim P_{\rm orb}^{\rm cl} $.
After that, the clump's angular momentum evolves on a timescale of $t_ {\rm evo} $
under the influence of the gravitational perturbation from the remnant planet.
Therefore, the time for the clump to travel to the NS can be approximated
as $ t_{\rm trav} \sim 2t_{\rm ret} + t_{\rm evo}$.
For the configurations with $ t_{\rm evo} \lesssim P_{\rm orb}^{\rm cl} $,
we have $2P_{\rm orb}^{\rm cl} < t_{\rm trav} \lesssim 3P_{\rm orb}^{\rm cl}$,
which means that the clumps will fall onto the NS within $\sim$ 2 --- 3
orbital periods after their birth.

\section{Discussions}\label{sec:factors}

In a binary system, the orbit will evolve due to various effects such as
mass loss, gravitational wave (GW) radiation, tidal dissipation, and magnetic
interactions. In this section, we discuss the effects of these evolutionary
channels on our planetary systems.

\begin{enumerate}
	\item Mass loss
	
	In our framework, the planet passes through the periastron repeatedly, and thus
	may experience multiple disruptions. Mass loss is a significant feature in
	the process, the effect of which on the orbit of the planet needs to be addressed.
	Especially, it should be clarified whether the planet still remains bound to the
	NS after multiple disruptions. During a partial disruption, the total mass loss is
	$ \Delta m = \Delta m_{\rm 1} + \Delta m_{\rm 2} $,
	where $ \Delta m_{\rm 1} $ and $ \Delta m_{\rm 2} $
	are the mass loss from the Lagrangian points $ L_{1} $ and $ L_{2} $,
	respectively. Usually, the mass loss is asymmetric with
	$ \Delta m_{\rm 1} > \Delta m_{\rm 2} $. The surviving core thus
	could obtain a kick velocity due to the asymmetry, which is mainly
	determined by the mass difference defined
	as $\Delta m_{\rm 12} = \Delta m_{\rm 1} - \Delta m_{\rm 2}$. If the kick
	velocity is too large, then the surviving core could be scattered away from the NS.
	Similar processes have been extensively studied by many authors, but mainly on
	the tidal disruption of gaseous giant planets \citep{Guillochon2011ApJ,Liu2013ApJ} and main
	sequence stars \citep{Manukian2013ApJ,Gafton2015MNRAS,Zhong2022ApJ}.
	It is found that the mass difference ($\Delta m_{\rm 12}$) is very sensitive
	to $r_{\rm p}$. It generally decreases with the increase of $r_{\rm p}$.
	In our case, we have $r_{\rm p} \gtrsim 2 r_{\rm td}$ ($ \beta \lesssim 0.5$),
	then the mass difference of $\Delta m_{\rm 12}$ is essentially very
	small even for gaseous planets \citep{Liu2013ApJ}. Additionally, the
	planet is a rocky one, which further markedly reduces the mass difference.
	The impulse kick caused by such a $\Delta m_{\rm 12}$ is small and will
	not affect the orbit of the surviving core. As a result, it remains to be
	bound to the NS and continues to revolve around the host.
	
	\item GWs and Tides
	
	The planet's orbit may be affected by gravitational wave
	radiation and tidal dissipation. Here we consider these two
	factors. For the GW emission, we have calculated the decay time of
	the orbit ($t_{\rm gw}$) by using equation (6) of \citet{Vick2017MNRAS}.
	It is found that the decay time is very long, $t_{\rm gw} > 10^{15}$ yr.
	For tidal dissipation, the decay time is even longer than the
	gravitational timescale \citep{Vick2017MNRAS}. So, in our cases,
	the effects of both tidal dissipation and GW radiation can be neglected.

	\item Alfv\'{e}n-wave Drag
	
	It should be noted that an object moving around an NS could interact
	with the magnetosphere of the NS when the orbit radius is too small,
	producing some Alfv\'{e}n wing structures. Such interactions will
	help the NS capture the object \citep{Cordes2008ApJ,Mottez2011AA}.
	However, in our cases, the periastron is still relatively further away
	from the NS. As a result, for the clumps that are of the size of a
	few kilometers, the orbit decay time is very long \citep[e.g.][]{Zhang2021ApJ}.
	Therefore, the Alfv\'{e}n-wave drag is not a dominant mechanism for
	the orbit decay of the planet and clumps.
	
\end{enumerate}

In short, the effects of mass loss, GWs, tides,
and magnetic fields on the evolution of the planet and clumps are
negligible in our framework.


\section{Conclusions} \label{sec:conclusion}

In this study, we concentrate on the systems in which a rocky planet
revolves around an NS in a highly elliptic orbit. The periastron of
the planet is close to the NS, $r_{\rm p} \sim 2r_{\rm td}$, so that
it will be partially disrupted every time it passes through the
periastron, forming a multi-body system. The dynamical evolution of
the clumps stripped off from the planet is investigated, taking
into account the joint gravitational interactions of the NS and the
surviving portion of the planet. The interactions among clumps
are neglected for simplicity. As a result, the NS, the remnant planet,
and a particular clump form a triple system. For the evolution of the
clump in such a system, our conclusions are as follows:
\begin{itemize}
	\item Gravitational perturbation from the remnant planet is a
	dominant factor that governs the evolution of the clump's orbit.
	Due to the perturbation, the eccentricity of the clump orbit increases
	quickly so that its specific angular momentum decreases to almost zero
	on a timescale of several orbital periods, causing the clump
	to effectively collide with the NS.
	\item The detailed evolution of the clumps is significantly affected by
	the orbital parameters and the structure of the planet.
	We have examined three types of planets:
	pure Fe, pure MgSiO$_{3}$, and two-layer planets (Fe core and MgSiO$_{3}$
	mantle). By assuming three typical orbital periods for the planet, i.e. 10 days,
	100 days, and 1000 days, it is found that the evolutionary timescale
	of the clump's angular momentum decreases with the increase
	of the orbital period and the planet's mass. For planets with different
	compositions, the trend of change is similar for the evolution time timescale,
	but the exact values differ significantly.
\end{itemize}

Generally, we see that the clumps will collide with the NS shortly after
their birth in a partial disruption event. The collision may lead to some
kinds of electromagnetic transient events. In fact, it has been
argued that the accretion of a small body by an NS can produce some
special kinds of gamma-ray bursts \citep{Colgate1981ApJ}.
GRB 101225A may occur in this way \citep{Campana2011Natur}.
The collision of an asteroid with an NS may also account for
X-ray bursts \citep{Huang2014,Geng2015ApJ_b,Dai2020,Geng2020} and/or fast radio
bursts \citep{Geng2015ApJ_b,Dai2016,Smallwood2019,Dai2020,Geng2020,Dai2020a}.
However, the exact phenomenon associated with the collision will depend on
many detailed factors, such as the composition and size of the clump, the magnetic
field and spin period of the NS, the impact parameter, etc. It may even depend on
the internal composition of the so-called neutron star, which may actually be
a strange quark star \citep{Geng2021Innov,Nurmamat2022arXiv221112026N}.
A detailed case study on the various possible transients associated with the
collision is still being conducted. Additionally, it should be noted that
the partial tidal disruption of a rocky planet near the periastron
is a very complicated process. In this study, we have taking some simplified
assumptions on the size and shape of the clumps. To clarify these issues,
numerical simulations on the tidal disruption process should be necessary,
which is beyond the scope of this study.

\section{Acknowledgments}

We would like to thank the anonymous referee for helpful
suggestions that led to significant improvement of our study
This work is supported by
the Chinese Academy of Sciences (CAS) ``Light of West China'' Program (No. 2019-XBQNXZ-B-016),
the Natural Science Foundation of Xinjiang Uygur Autonomous Region (No. 2022D01A363),
the Major Science and Technology Program of Xinjiang Uygur Autonomous Region (Nos. 2022A03013-1, 2022A03013-3),
the National Natural Science Foundation of China (Grant Nos.
12041304, 12033001, 12273028, 12233002, 12041306, 12147103),
the Youth Innovations and Talents Project of Shandong Provincial Colleges and Universities (Grant No. 201909118),
National Key R\&D Program of China (2021YFA0718500),
National SKA Program of China No. 2020SKA0120300,
the special research assistance project of the CAS,
and by the Operation, Maintenance and Upgrading Fund for Astronomical Telescopes and
Facility Instruments, budgeted from the Ministry of Finance of China (MOF) and administrated by the CAS.

\section*{Data Availability}


The data underlying this article are available from the corresponding author upon reasonable request.




\bibliographystyle{mnras}
\bibliography{reference} 








\bsp	
\label{lastpage}
\end{document}